\documentclass[fleqn,usenatbib]{mnras}
\usepackage[T1]{fontenc}
\let\VANthebibliography\thebibliography
\def\thebibliography{\DeclareRobustCommand{\VAN}[3]{##3}\VANthebibliography}

\usepackage{xcolor,graphicx,xspace,color,longtable}
\usepackage{amssymb,amsmath,shadow,bezier,curves,rotating}
\usepackage{graphicx,chngcntr}
\graphicspath{ {./images/}}

\newcommand {\h}    {$h^{-1}\mathrm{Mpc}$}

\newcommand {\hm}   {$h^{-1}M_{\odot}$}

\newcommand {\Gal}  {$\mathrm{GalWCat}$} 
\newcommand {\G}    {$\mathrm{GLAM}$}
\newcommand {\U}   {$\mathrm{Uchuu}$}

\title[Cluster Correlation Function]{The Correlation Function and Detection of Baryon Acoustic Oscillation Peak from the  Spectroscopic $\mathtt{SDSS-GalWCat}$ Galaxy Cluster Catalogue}

\author[M. H. Abdullah]{Mohamed H. Abdullah,$^{1,2}$\thanks{E-mail: melha004@ucr.edu} Anatoly Klypin,$^{3,4}$ Francisco Prada,$^{5}$ Gillian Wilson,$^{6}$  
\newauthor
Tomoaki Ishiyama,$^{1}$ Julia Ereza,$^{5}$
\\
\scriptsize
$^{1}$ Digital transformation enhancement council, Chiba University, 1-33, Yayoi-cho, Inage-ku, Chiba, 263-8522, Japan\\
\scriptsize $^{2}$ Department of Astronomy, National Research Institute of Astronomy and Geophysics, Cairo, 11421, Egypt\\
\scriptsize $^{3}$ Astronomy Department, New Mexico State University, Las Cruces, NM 88001\\
\scriptsize $^{4}$ Department of Astronomy, University  of Virginia, Charlottesville, VA 22904, USA\\
\scriptsize $^{5}$ Instituto de Astrof\'isica de Andaluc\'ia (CSIC), Glorieta de la Astronom\'ia, E-18080 Granada, Spain\\
\scriptsize $^6$ Department of Physics, University of California Merced, 5200 Lake Rd., Merced, CA 95343, USA
}

\begin{document}
\label{firstpage}
\pagerange{\pageref{firstpage}--\pageref{lastpage}}
\maketitle

\begin{abstract}
We measure the two point correlation function (CF) of 1357 galaxy clusters with a mass of $\log_{10}{M_{200}}\geq 13.6$~\hm~and at a redshift of $z \leq 0.125$.
This work differs from previous analyses in that it utilizes a spectroscopic cluster catalogue, $\mathtt{SDSS-GalWCat}$, 
to measure the CF and detect the baryon acoustic oscillation (BAO) signal.
Unlike previous studies which use statistical techniques, we compute covariance errors directly by generating a set of 1086 galaxy cluster lightcones from the GLAM $N$-body simulation.
Fitting the CF with a power-law model of the form $\xi(s) = (s/s_0)^{-\gamma}$, we determine the best-fit correlation length and power-law index at three mass thresholds. 
We find that the correlation length increases with increasing the mass threshold while the power-law index is almost constant. 
For $\log_{10}{M_{200}}\geq 13.6$~\hm, we find $s_0 = 14.54\pm0.87$~\h~and $\gamma=1.97\pm0.11$.
We detect the BAO signal at $s = 100$~\h~with a significance of $1.60 \sigma$. Fitting the CF with a $\Lambda$CDM model, we find $D_\mathrm{V}(z = 0.089)\mathrm{r}^{fid}_d/\mathrm{r}_d = 267.62 \pm 26$ \h, consistent with Planck 2015 cosmology. We present a set of 108 high-fidelity simulated galaxy cluster lightcones from the high-resolution \U~N-body simulation, employed for methodological validation. We find 
$D_\mathrm{V}(z = 0.089)/r_d = 2.666 \pm 0.129$, indicating that our method does not introduce any bias in the
parameter estimation for this small sample of galaxy clusters.
\end{abstract}

\vspace{-0.75cm}
\section{Introduction}
The investigation of galaxy clusters and their clustering properties provides insights into understanding of the large-scale structure of the Universe \citep{Peebles80, Moscardini01, Allen11}. 
The CF is a key tool for quantifying the clustering properties of galaxy clusters \citep{Peebles74,Klypin94}. Analyzing the CF is particularly crucial for constraining models of galaxy cluster formation and evolution, which still pose significant challenges in our understanding of the process \citep[e.g.,][]{Voit05, Kravtsov12}. Additionally, the CF plays a vital role in testing the fidelity of numerical simulations, including MultiDark $N$-body simulations \citep{Klypin16} and \U~$N$-body simulations \citep{Ishiyama21}.

\begin{figure*}\hspace{0cm}
\centering
\includegraphics[width=1\linewidth]{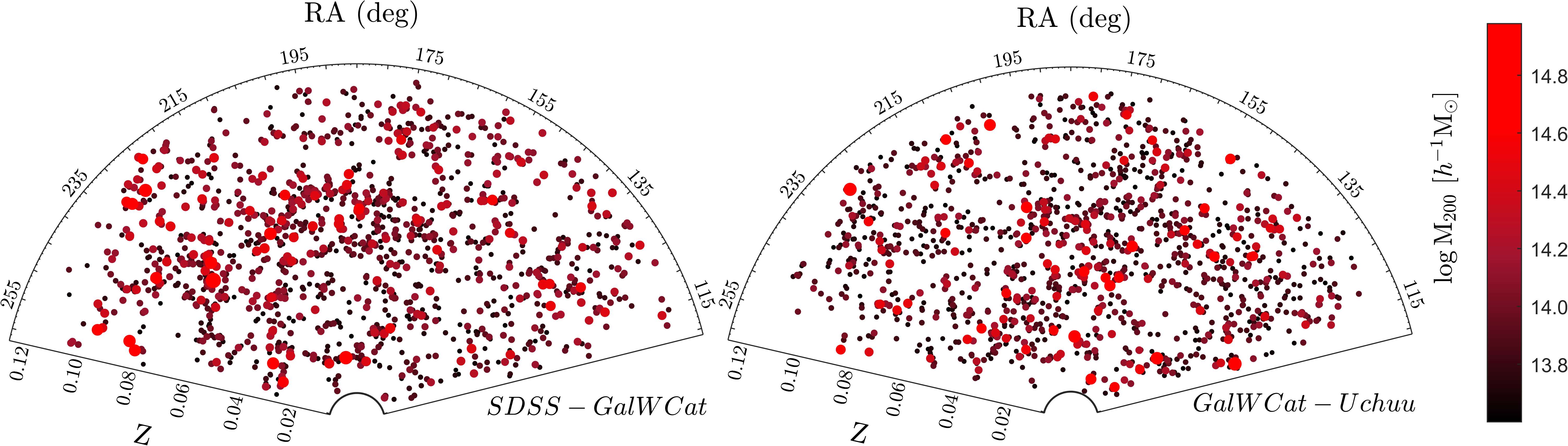} \vspace{0cm}
    \caption{
Slice through SDSS-\Gal~(observations; left), and one of the \Gal-\U~lightcones (simulation; right). 
The lightcones are centered at ($\alpha$,~$\delta$) = $(185^\circ$,~$0^\circ$) with width of $\delta = 90^\circ$. The color and size of the circles in each case indicates the mass of the cluster. 
}
\label{fig:Fig01}
\end{figure*}

On small scales of a few Mpcs, the CF is frequently characterized by a power-law. The slope of this power-law holds significant implications for the underlying matter density and the structure growth rates in the Universe. Additionally, the amplitude of the CF serves as a valuable indicator of the bias exhibited by the galaxy cluster population \citep[e.g.,][]{Colberg00, Moscardini00, Collins00, Bahcall04, Desjacques18}. 

On large scales, the CF exhibits the BAO signal \citep{cole05,Eisenstein05}. The BAO signal provides a powerful tool for measuring the expansion history of the Universe and the growth of cosmic structure (e.g., \citealp{Sanchez14,Alam21}). Remarkably, the BAO signal has been successfully detected in the CF of galaxy clusters at a scale of $\sim 100$~\h, employing various cluster samples. For instance, \citet{Estrada09} detected the BAO signal using the maxBCG cluster sample \citep{Koester07}. Additionally, \citet{Hong12}, \citet{Veropalumbo14}, and \citet{Hong16} detected the BAO signal using cluster samples obtained from \citet{Wen09}, \citet{Wen12}, and \citet{Wen15}, respectively.

In this paper, we present a new measurement of the galaxy cluster CF and the cosmological distance $D_\mathrm{V}$ at redshift $z =0.089$ by utilizing the spectroscopic SDSS-$\mathtt{GalWCat}$\footnote{\url{http://cdsarc.u-strasbg.fr/viz-bin/cat/J/ApJS/246/2}} galaxy cluster catalogue \citep{Abdullah20a}. We compute the covariance matrix directly using a set of 1086 independent mock catalogues extracted from GLAM\footnote{\url{http://skiesanduniverses.iaa.es/Simulations/GLAM/}} N-body simulations \citep{Klypin18}. Additionally, we present a set of 108 high-fidelity simulated galaxy cluster lightcones from the Uchuu\footnote{\url{http://skiesanduniverses.org/Simulations/Uchuu/}} N-body simulation \citep{Ishiyama21}. 
These lightcones are used for verifying the accuracy and reliability of our approach. Furthermore, in prospective applications, the employment of lightcones extends towards exploring cluster internal dynamics and scaling relationships, considering the projection effects.


The advantages of using the $\mathtt{GalWCat}$ cluster catalog are as follows. Clusters were identified by the well-known Finger-of-God effect (see, \citealp{Jackson72,Kaiser87,Abdullah13}). Cluster membership was determined by applying the GalWeight technique, which has been shown to have $>98\%$ accuracy in correctly assigning cluster membership \citep{Abdullah18}. A cluster mass was calculated for each cluster individually using the dynamics of member galaxies via the virial theorem (e.g., \citealp{Limber60,Abdullah11}), and then corrected for the surface pressure term (e.g., \citealp{The86,Carlberg97}).

The paper is organized as follows. Section~\ref{sec:data} provides an overview of the $\mathtt{GalWCat}$ cluster catalogue, as well as details of both the \U~and \G~simulations that we utilize throughout. In Section~\ref{sec:CF}, we present the CF and measure the power-law clustering as a function of different cluster mass bins. Section~\ref{sec:bao} delves into the analysis of the BAO signal and discusses its cosmological implications. Finally, we summarize our results and conclusions in Section~\ref{sec:conc}. 
Throughout the paper, we assume a $\Lambda$CDM cosmology based on Planck 2015 results \citep{Planck15}, i.e., adopting the following values of cosmological parameters: $\Omega_M=0.3089$, $\Omega_\Lambda=0.6911$, $\Omega_b=0.0486$, and $h=0.6774$. 
Note that we employ the term 
`log' to mean log base-10 (i.e., $\log_{10}$).

\vspace{-0.5cm}
\section{Data and simulation lightcones} \label{sec:data}
\subsection{The $\mathtt{SDSS-GalWCat}$} \label{sec:cat}
From the SDSS-DR13 spectroscopic dataset, \citet{Abdullah20a} constructed a publicly available catalogue of 1800 galaxy clusters, $\mathtt{GalWCat}$.  
The mass of each cluster, $M_{200}$, was estimated by applying the virial theorem at the radius within which the density, $\rho$, satisfied the condition $\rho=200\rho_c$, where $\rho_c$ is the critical density of the Universe (e.g.,  \citealp{Carlberg97,Klypin16}). The 1800 clusters in the catalogue have redshifts in the range $0.01 < z < 0.2$ (with a mean redshift of $z = 0.089$) and masses in the range $13.6 < \log{M_{200}} < 15.1$~\hm.

The selection function of $\mathtt{GalWCat19}$ is $\mathcal{S}_z (D) = 1.07 \exp\left[{-\left(\frac{D}{293.4}\right)^{2.97}}\right]$, where $D$ is the comoving distance to the cluster \citep{Abdullah20b}. To mitigate shot noise in the correlation function and the BAO signal, we limit our analysis to $\mathcal{S}_z \lesssim 0.2$ at  maximum redshift of $z \leq 0.125$. Additionally, we exclude the clusters in the three small separated stripes in the SDSS sky to eliminate selection effect of the boundary on the CF and the BAO signal.
The resulting subsample, which we refer to as SDSS-GalWCat, consists of 1357 clusters. Throughout the remainder of this paper, we utilize the SDSS-GalWCat catalog to calculate the correlation function and detect the BAO signal.

\begin{figure*}\hspace{0cm}
\centering
    \includegraphics[width=0.9\linewidth]{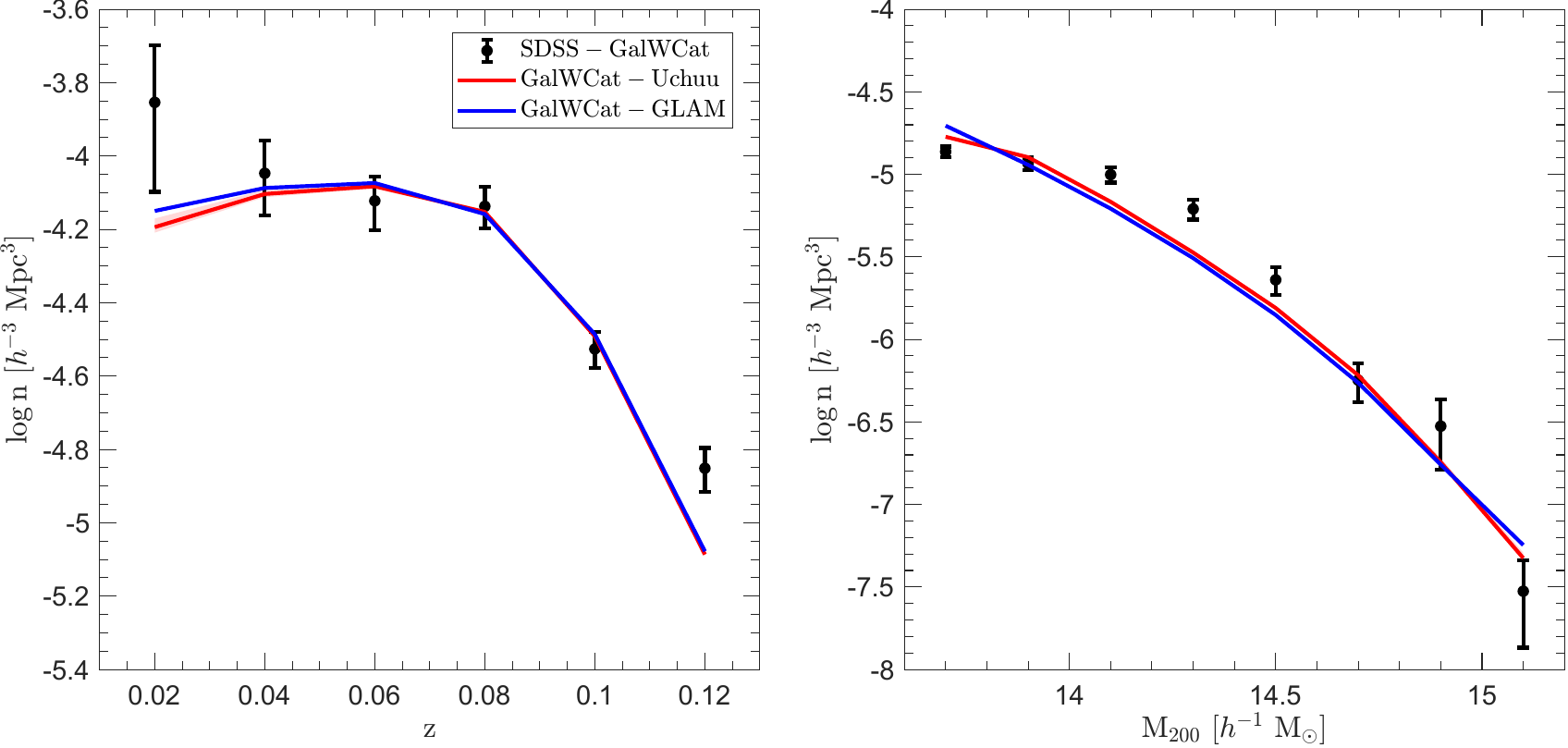} \vspace{-0.25cm}
    \caption{Redshift (left) and mass (right) comoving number densities of SDSS-\Gal~clusters and the averages of 108 \Gal-\U~and 1086 Gal-\G~lightcones. The uncertainties show $1\sigma$ scatter calculated from the covariance matrix which is derived from the \Gal-\G~lightcones.}
    \label{fig:Fig02}
\end{figure*}

\vspace{-0.4cm}
\subsection{Cosmological simulations} \label{sec:sim}
We generate high-fidelity simulated galaxy cluster lightcones (see e.g. \citealp{Bernyk16,Abdullah20a} for the description of lightcone construction) with the same selection function and footprint as SDSS-GalWCat utilizing the \U~simulation. The redshift-space distortion is taken into account. These lightcones are crucial for our methodological validation, where we systematically evaluate the precision of calculating the CF and detecting the BAO using a relatively small sample of 1357 clusters. This analysis aims to uncover any potential biases in parameter estimation that may be introduced by our methodology. Additionally, these simulated lightcones are generated for future and external applications. They will be valuable for exploring cluster scaling relationships and internal dynamics, taking into account the influence of projection effects.

The \U~simulation is selected from the \U~suite of large, high-resolution $N$-body simulations \citep{Ishiyama21}, which were carried out assuming Planck 2015 cosmology \citep[][]{Planck15}.  
\U~is a cosmological $N$-body simulation of $12800^3$ particles in a box of comoving side-length of 2000~\h. 
Particle mass resolution and gravitational softening length are $3.27 \times 10^8$~\hm\ and $4.27~h^{-1}\mathrm{kpc}$, respectively. \U~was created using the massively parallel $N$-body TreePM code, \textsc{greem} \citep{Ishiyama09,Ishiyama12}. Haloes and subhaloes were identified with \textsc{rockstar} \citep{Behroozi13a} and merger trees constructed with \textsc{consistent trees} \citep{Behroozi13b}. 
In order to be consistent with how the $\mathtt{GalWCat}$ cluster masses are calculated, we use $M_{200}$ for Uchuu cluster masses. Here, we analyze the snapshot at redshift $z\sim 0.09$. We divide the 2000~\h~box into $108$ smaller boxes to generate $108$ independent lightcones that replicate the same footprint and redshift interval as SDSS-\Gal~and assume the line-of-sight is along the $z$-direction. We refer to the lightcones generated from \U~as~\Gal-\U. The \Gal-\U~as well as \Gal-\G~(see Section \ref{sec:covmat}) are available at \url{http://skun.iaa.es/SUsimulations/UchuuDR3/}.

Fig.~\ref{fig:Fig01} shows a slice through the lightcone for SDSS-\Gal~(left) and through one of the 108 lightcones from \Gal-\U~(right), with the mass of each cluster indicated by its color. The figure demonstrates the similarities between the lightcone of SDSS-\Gal~and that constructed from \Gal-\U. Fig.~\ref{fig:Fig02} shows the comoving number density (left) and the mass density (right) of SDSS-\Gal~clusters (black circles) and the averages of 1086 \Gal-\G~(blue lines) and  108 \Gal-\U~(red line) lightcones. There is good agreement in the cluster number density and the mass density between the SDSS-\Gal~and both \Gal-\G~and \Gal-\U~lightcones.

\subsection{The covariance matrix} \label{sec:covmat}
The covariance matrix is a crucial ingredient for clustering analyses. It measures the correlation between correlation function bins. In our analysis, we utilize the \G~simulation \citep{Klypin18} to calculate the covariance matrix and the uncertainties of the CF. \G~is a cosmological $N$-body simulation of $2000^3$ particles in a box of comoving length 1000~\h, particle mass resolution of $1.07\times 10^8$~\hm, and gravitational softening length of $250~h^{-1}\mathrm{kpc}$. It assumes a standard $\Lambda$CDM cosmology with $\Omega_M=0.307$, $\Omega_\Lambda=0.693$, $\Omega_b=0.0483$, $\sigma_8 = 0.8288$, and $h=0.6777$.
The distinct haloes are identified with the Bound Density Maximum halo finder \citep{Klypin97}. We generate $1086$ independent lightcones from $1086$ realizations of a snapshot at redshift $z\sim0.11$. The lightcones are generated with the same selection function and the sky footprint as SDSS-GalWCat.

The covariance matrix is calculated from the 1086 independent \G~lightcones as
\begin{equation}
\label{eq:err}
C_{ij}=\frac{1}{N-1}\sum_{l=1}^{N}\left(\xi^{l}_{i}-\overline{\xi}_{i}\right)\left(\xi^{l}_{j}-\overline{\xi}_{j}\right),
\end{equation}
\noindent where $N$ is the number of lightcones, $\xi^{l}_{i}$ is the CF of the $l^{th}$ lightcone at the $i^{th}$ bin, and $\overline{\xi}_{i}$ represents the mean value of all lightcones at the $i^{th}$ bin. The error bars of $\xi(s)$ are given by the diagonal elements as $\sigma_{i}=\sqrt{C_{ii}}$. We address the comparison between covariance matrices derived from \G~lightcones and those obtained from jackknife and bootstrap resampling methods in Appendix~\ref{App:App1} and \ref{fig:Cov}.

From the covariance matrix, we define the $\chi^2$ statistic as
\begin{equation} \label{eq:chi}
    \chi^2 = \frac{N-p-2}{N-1} (\xi^{\text{Data}}-\xi^{\text{Model}})C^{-1}(\xi^{\text{Data}}-\xi^{\text{Model}})^T,
\end{equation}
\noindent where $p$ is the number of degrees-of-freedom and $N$ is the number of GLAM lightcones, including the Hartlap correction \citep{Hartlap07}. Throughout the paper we apply the Monte Carlo Markov Chain (MCMC) technique using the full covariance matrix. We adopt the standard Gaussian likelihood where the likelihood $\mathcal{L} \propto\exp{\left(\frac{-\chi^2}{2}\right)}$.

\vspace{-0.5cm}
\section{Cluster Correlation Function} \label{sec:CF}
In this section, we present the CFs derived from SDSS-\Gal~as a function of cluster mass, and compare them with those obtained from the \Gal-\U~lightcones. We use the Landy–Szalay estimator \citep{Landy93} to calculate the CF, $\xi(s)$:
\begin{equation}
\label{eq:LS}
\xi(s)=\left[DD(s)\frac{N_{RR}}{N_{DD}}-2\;DR(s)\frac{N_{RR}}{N_{DR}}+RR(s)\right]/RR(s),
\end{equation}
\noindent where $DD(s)$, $DR(s)$, and $RR(s)$ are the number of data-data, data-random, and random-random pairs within a separation annulus of $s\pm\Delta s/2$, and $N_{DD}$, $N_{DR}$, and $N_{RR}$ are the normalization factors of the pairs. We generate random samples based on the sky footprint and redshift selection function of SDSS-\Gal~(see \S~\ref{sec:cat}). 
The number of objects generated in each random sample is $\sim25$ times larger than that of the SDSS-\Gal.

\begin{figure}\hspace{0cm}
\centering
\includegraphics[width=0.95\linewidth]{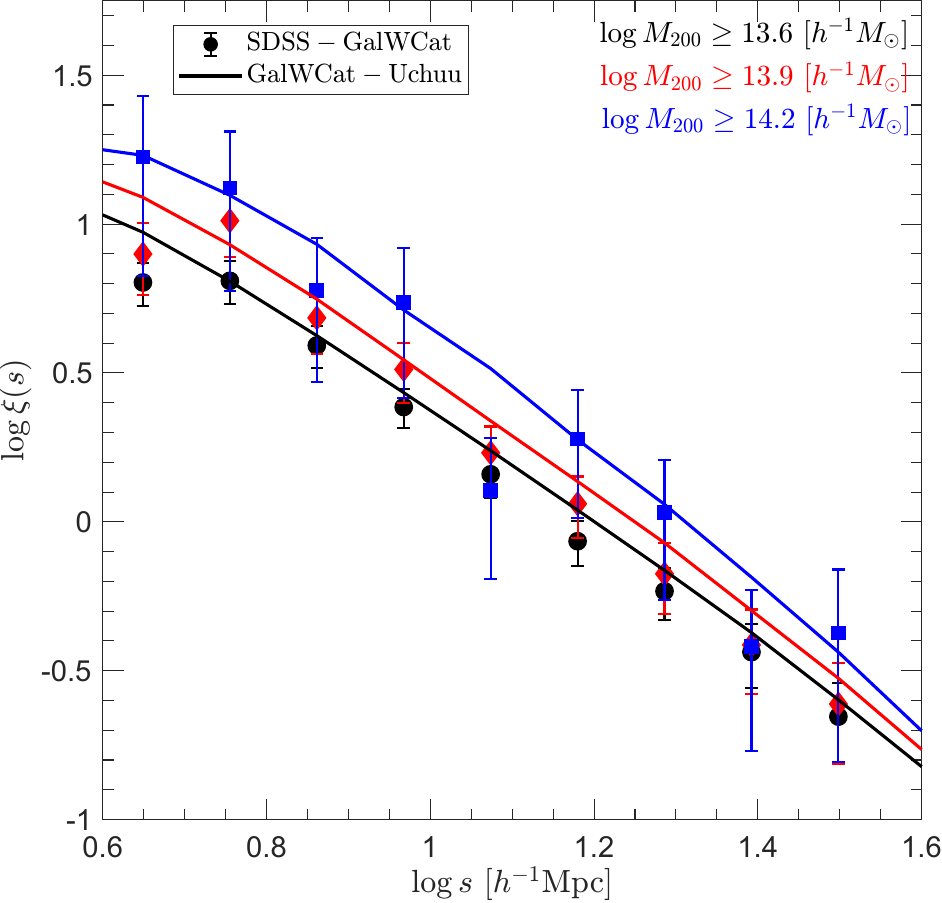} \vspace{-0.25cm}
    \caption{
    Comparison between the correlation function, $\xi(s)$, derived from observations (colored symbols) and simulations (colored lines), as a function of the separation, for three different mass thresholds (see \S~\ref{sec:CF}). The $1\sigma$ uncertainties were calculated from the \G~lightcones.
    }
    \label{fig:Fig03}
\end{figure}

On small scales, the CF can be fit by a power-law of the form
\begin{equation} \label{pwoer}
\xi(s)=\left(\frac{s}{s_0}\right)^{-\gamma} ,
\end{equation}
\noindent where $s_0$ is the correlation length and $\gamma$ is the slope. The value of the correlation length is mass dependent, and has been shown to increase with cluster mass (e.g., \citealp{ Bahcall04,Croft97,Estrada09}). 
We perform the $\chi^2$ fit on scales of $\log{5} < \log{s} < \log{40}$~[\h], using separation bins of  width $\log{\Delta s} = 0.1062$~[\h].
Fig.~\ref{fig:Fig03} presents the SDSS-\Gal~CFs for three different cluster mass thresholds ([13.6, 13.9, 14.2]~\hm), and compares them to those obtained from the 108 \Gal-\U\ lightcones. We note the very close agreement between the observations and simulations. 

In Table~\ref{tab:[[Fit]]} we show the best fit correlation length and power-law slope derived for either the SDSS-\Gal~and \Gal-\U~for each of the three mass thresholds. As expected, we find that $\gamma$, the slope of the power law, is almost constant with cluster mass while the correlation length, $s_0$, increases with increasing cluster mass. This result suggests a strong connection between the mass of galaxy clusters and their spatial distribution patterns. More massive clusters exert more gravitational pull, resulting in stronger clustering. Also, as expected, we find that the value of $s_0$ is considerably larger for clusters than for galaxies, e.g., $s_0 = 5.91$~\h~for SDSS galaxies \citep{Zehavi04} and $s_0 = 5.05$~\h~for 2dF galaxies \citep{Hawkins03}. 

\begin{table}
	\centering
	\caption{Best-fit correlation function parameters.
            L to R, columns show sample utilized (observations or simulations), mass limit (in units of~\hm), correlation length (in~\h), and slope.}
	\label{tab:[[Fit]]}
	\begin{tabular}{cccc}
	\hline
    Sample & $\log{M_{200}}$ & $s_0$ & $\gamma$\\
    & [\hm] & (\h) &\\
		\hline
SDSS-\Gal   &$\geq 13.6$ & $14.54\pm0.87$ & $1.97\pm0.11$ \\
SDSS-\Gal   &$\geq 13.9$ & $16.05\pm1.29$ & $2.15\pm0.19$ \\
SDSS-\Gal   &$\geq 14.2$ & $18.87\pm2.48$ & $2.05\pm0.32$ \\
   \hline
\Gal-\U   &$\geq 13.6$ & $15.70\pm0.04$ & $1.87\pm0.01$ \\
\Gal-\U   &$\geq 13.9$ & $17.45\pm1.05$ & $1.95\pm0.01$ \\
\Gal-\U   &$\geq 14.2$ & $20.10\pm2.09$ & $2.10\pm0.01$ \\
   \hline
	\end{tabular}
\end{table}

\begin{figure*}\hspace{0cm}
\centering
\includegraphics[width=1\linewidth]{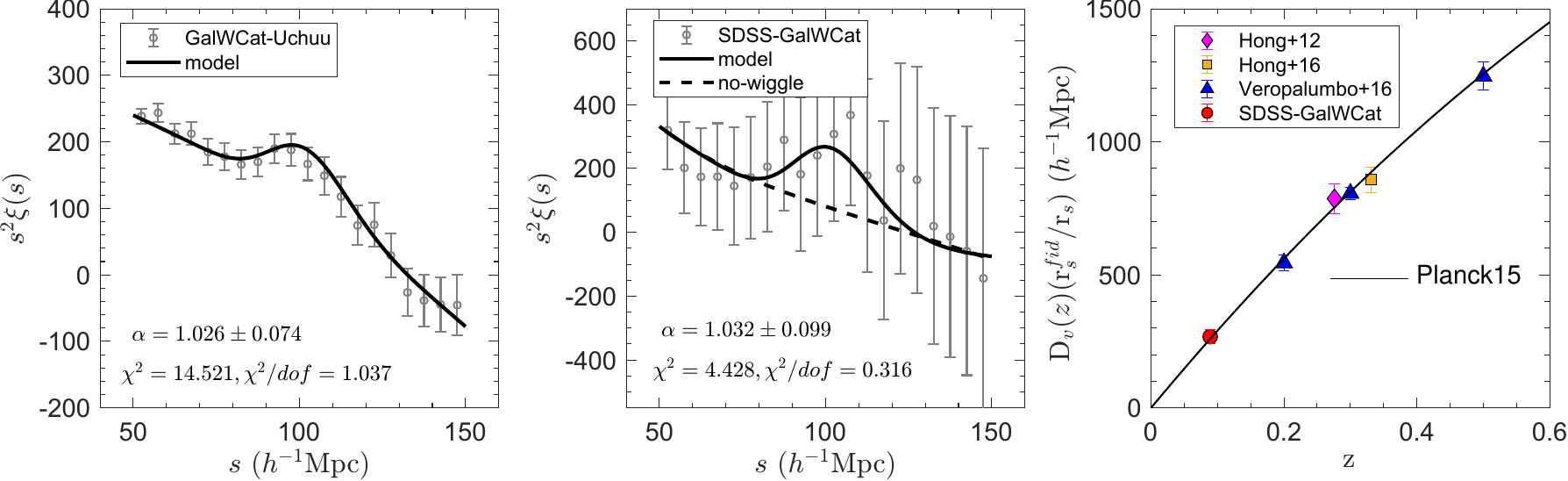} \vspace{-0.25cm}
    \caption{Measured (grey symbols) and best-fit (black lines) CFs from the \Gal-\U~(left panel) simulated lightcones, and the SDSS observations (middle panel) at z = 0.089 using a bin width of  $\Delta s = 5$~\h. The uncertainties in both panels are obtained from the GLAM lightcones (for Uchuu it is $1\sigma$ divided by the square-root of the 108 lightcones). 
    The dashed line in the middle panel is the best fit no-wiggle model ($\chi^2 = 6.878$), used to calculate the significance of the BAO signal. 
    The distance $\mathrm{D}_V(z)/(\mathrm{r}^{fid}_s/rs)$–redshift relation is presented in the right panel. Our measurement at $z=0.089$ is represented by the red point, while the other points correspond to distance measurements derived from other studies on the CF of galaxy clusters. The black curve represents the prediction from the $\Lambda$CDM model based on the Planck 2015 cosmology.}
    \label{fig:Fig04}
\end{figure*}

\vspace{-0.5cm}
\section{BAO detection and cosmology} \label{sec:bao}
In this section, we analyze the BAO signal within the framework of a $\Lambda$CDM model with \citet{Planck15} cosmology. The methodology we use to measure the BAO positions is
adapted from \citet{Ross15} (see also e.g. \citealp{Veropalumbo16,Paez22}).
The typical approach is to model the nonlinear power spectrum $P_{\mathrm{nl}}$ as (e.g., \citealp{Ross15})
\begin{equation}
P_{\mathrm{nl}}(k) = \left[P_{\mathrm{lin}}(k) -
P_{\mathrm{nw}}(k)\right]\exp \left(\frac{-k^2
  \Sigma_{\mathrm{nl}}^2}{2}\right)+P_{\mathrm{nw}}(k),
\end{equation}
where $P_{\mathrm{lin}}$ and $P_{\mathrm{nw}}$ are the linear and no-wiggle power spectra and $\Sigma_{\mathrm{nl}}$ is a parameter modeling the non-linear degradation (e.g., \citealp{Estrada09,Xu12}). $P_{\mathrm{lin}}$ and $P_{\mathrm{nw}}$ are computed at $z =0.089$ using \textsc{Colossus} python package \citep{Diemer18}. We adopt $\Sigma_{\mathrm{nl}} = 8$~\h~in our analysis (e.g. \citealp{Hong12,Hong16}). 
The template correlation function with damped BAO is then given as
\begin{equation}
\xi_{\mathrm{temp}}(s) = \frac{1}{2\pi}\int{P_{\mathrm{nl}}(k)\frac{\sin{(ks)}}{ks}k^2dk} 
\end{equation}
Finally, we fit the cluster CF with the model
\begin{equation}
\xi_{\mathrm{model}}(s) = B_{\xi}^2\xi_{\mathrm{temp}}(\alpha s)+\frac{a_1}{s^2}+\frac{a_2}{s}+a_3,
\end{equation}
where $B_{\xi}$ is a multiplicative constant allowing for an unknown large-scale bias between the distribution of clusters and the underlying matter density field, $\alpha$ is a dilation cosmological parameter that can be estimated by comparing the data to the template, and  $a_1$, $a_2$ and $a_3$ are free parameters. Note that polynomial terms include effects caused by, e.g. errors made in the assumption of the model cosmology, scale-dependent bias, and redshift-space distortion (see e.g. \citealp{Ross15}).

The parameter $\alpha$ is related to physical distances via
\begin{equation}
\label{eq:alph}
\alpha=\frac{D_\mathrm{V}(z)r_\mathrm{d}^\mathrm{fid}}{D_\mathrm{V}^\mathrm{fid}(z)r_\mathrm{d}},
\end{equation}
and 
\begin{equation}
\label{eq:Dv}
D_\mathrm{V}(z) = \left[ (1+z)^2 D_\mathrm{A}(z)^2 \frac{cz}{H(z)}\right]^{1/3},
\end{equation}
where $H(z)$ is the Hubble parameter, $D_\mathrm{A}(z)$ is the comoving angular diameter distance, and $r_\mathrm{d}$ is the sound horizon at the baryon drag epoch, which can be accurately calculated for a given cosmology using CAMB package \citep{Lewis00}. For our fiducial cosmology, $D_\mathrm{V}^\mathrm{fid}(0.089) = 259.32$~\h~ and $r_\mathrm{d}^\mathrm{fid}=99.81$~\h.

We perform the $\chi^2$ fit (Eq. \ref{eq:chi}) on scales of $50 < s < 150$~\h, using separation bins of width $\Delta s = 5$~\h. 
Fig.~\ref{fig:Fig04} shows the measured and the best-fit model of the CF from the \Gal-\U~lightcones and the SDSS-\Gal~observations. We find $\alpha=1.026\pm0.074$ and $\alpha=1.032\pm0.099$ for \Gal-\U~and SDSS-\Gal, respectively. From our dilation parameter estimate for the observed CF we find $D_\mathrm{V}(z = 0.089)\mathrm{r}^{fid}_d/\mathrm{r}_d = 267.62 \pm 26$ \h~which is consistent with Planck 2015 cosmology. For the \Gal-\U~lightcone measurements we find $D_\mathrm{V}(z=0.089)/r_\mathrm{d} = 2.666 \pm 0.192$, indicating that our method does not introduce any bias in the parameter estimation. For the SDSS-\Gal~observations, the best-fit nonlinear BAO model gives $\chi^2 = 4.428$ while the no-wiggle model gives $\chi^2 = 6.878$, and is rejected at $1.56 \sigma$. The right panel of Fig.~\ref{fig:Fig04} presents the distance $\mathrm{D}_V(z)/(\mathrm{r}^{fid}_s/rs)$–redshift relation. The plot displays measurements of $\mathrm{D}_V(z)$ obtained from various studies of galaxy cluster CF across different redshifts. Our new measurement at z=0.089 complements and strengthens the previous studies, providing additional support for the consistency of observations with the Planck 2015 cosmology.

\vspace{-0.5cm}
\section{Conclusions} \label{sec:conc}
In this work, unlike previous analyses which utilized photometric galaxy cluster catalogs, we utilized the spectroscopic cluster catalogue, $\mathtt{SDSS-GalWCat}$, to measure the CF. $\mathtt{SDSS-GalWCat}$ contains 1357 galaxy clusters with a mass of $\log_{10}{M_{200}}\geq 13.6$~\hm~and at a redshift of $z \leq 0.125$. 
We also created a set of 108 galaxy cluster lightcones using \U~N-body simulations. There is good agreement in the cluster number density between the SDSS-GalWCat and GalWCat-Uchuu lightcones. 
In contrast to previous analysis which calculated CF uncertainties utilizing resampling methods, we computed covariance errors directly from a set of 1086 galaxy cluster lightcones which we created from the GLAM N-body simulations. We found that jackknife and bootstrap resampling methods yield noisier matrices in contrasting with the smoother lightcone-derived matrix. We also shown that both jackknife and bootstrap errors are larger and noisier than lightcone errors in the range $0\leq s \leq 200$ \h.

On small scales, we fitted the CF by a power-law of the form $\xi(s) = (s/s_0)^{-\gamma}$, and determined the best-fit correlation length and power-law index for three cluster mass thresholds (Table~\ref{tab:[[Fit]]}). We found good agreement between the correlation functions measured from observations and those derived from the \U~lightcones. We noted that, the correlation length increases with increasing the mass threshold while the power-law index is almost constant ($\sim 2$) for both the data and the simulation.
For $\log_{10}{M_{200}}\geq 13.6$~\hm~we determined $s_0 = 14.54\pm0.85$~\h~and $\gamma=1.97\pm0.11$ for the data. For \U~lightcones, we found that $s_0$ is about 1 \h~larger than that of SDSS-\Gal ~at each mass threshold.

On large scales, we detected the peak of baryon acoustic oscillations (BAO) in the CF at $s = 100$~\h~ with a significance of $1.60 \sigma$. By fitting the observed CF with a $\Lambda$CDM model, we found $\alpha=1.032\pm0.099$ and $D_\mathrm{V}(z = 0.089)\mathrm{r}^{fid}_d/\mathrm{r}_d = 267.62 \pm 26$~\h. This is consistent with the fiducial cosmology obtained by the Planck 2015 data. 

\vspace{-0.5cm}
\section*{Acknowledgements}
FP and JE acknowledge financial support from the grant CEX2021-001131-S funded by MCIN/AEI/ 10.13039/501100011033.
TI has been supported by IAAR Research Support Program in Chiba University Japan, MEXT/JSPS KAKENHI (Grant Number JP19KK0344, JP21F51024, and JP21H01122), MEXT as ``Program for Promoting Researches on the Supercomputer Fugaku'' (JPMXP1020200109 and JPMXP1020230406), and JICFuS.
GW gratefully acknowledges support from the National Science Foundation through grant AST-2205189 and from HST program number GO-16300. Support for program number GO-16300 was provided by NASA through a grant from the Space Telescope Science Institute, which is operated by the Association of Universities for Research in Astronomy, Incorporated, under NASA contract NAS5-26555.

\section*{Data Availability}
The $\mathtt{GalWCat}$ galaxy cluster catalogue is available at \url{http://cdsarc.u-strasbg.fr/viz-bin/cat/J/ApJS/246/2}. The Uchuu simulation is available at
\url{http://skiesanduniverses.org/Simulations/Uchuu/} and the GLAM simulation is available at \url{https://www.skiesanduniverses.org/Simulations/GLAM/}. The \Gal-\U~and \Gal-\G~lightcones are available at \url{http://skun.iaa.es/SUsimulations/UchuuDR3/}.

\bibliographystyle{mnras}
\bibliography{ref}

\appendix
\vspace{-0.5cm}
\section{Comparison between covariance matrices} \label{App:App1}
\begin{figure*}\hspace{0cm}
\centering
\includegraphics[width=1\linewidth]{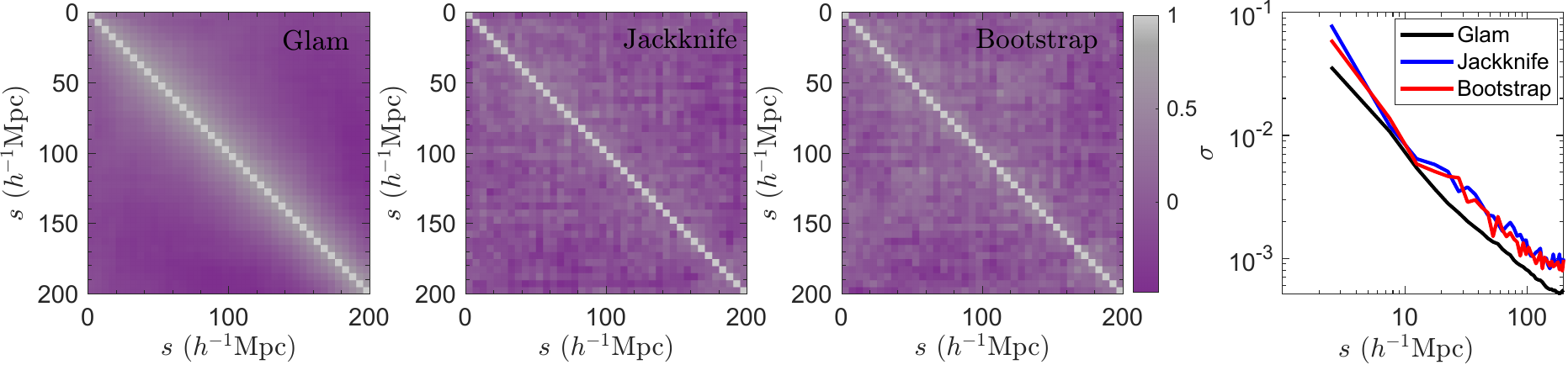} \vspace{-0.5cm}
    \caption{Correlation matrices $(C_{ij}/\sqrt{C_{ii}C_{jj})}$ derived from \G~lightcones, in addition to jackknife and bootstrap methods. The \G~lightcones provide the least scattered covariance matrix. The uncertainty given by the diagonal elements as $\sigma_{i}=\sqrt{C_{ii}}$ is shown in the right panel.
    }
    \label{fig:Cov}
\end{figure*}

To calculate the covariance matrix using the jackknife method, we utilize a lightcone from \G. We divide the celestial region covered by the lightcone into 48 evenly spaced subareas. 
By systematically excluding one subarea at a time, we create distinct subsamples for which we calculate the covariance matrix. 
For the bootstrap method, we perform 50 resampling iterations, drawing samples with replacement and calculating the covariance matrix 
from the resampled data sets.
In Figure \ref{fig:Cov}, we compare covariance matrices derived from \G~lightcones with those obtained from jackknife and bootstrap resampling methods. Resampling methods yield noisier matrices, contrasting with the smoother lightcone-derived matrix. 
Figure \ref{fig:Cov} (right panel) shows that both jackknife and bootstrap errors are larger and noisier than lightcone errors across all bins. This underscores the impact of resampling on stability and precision in covariance estimates, emphasizing the need for careful interpretation in numerical simulations.

\bsp
\label{lastpage}
\end{document}